\documentclass[journal, onecolumn]{IEEEtran}
\usepackage{graphicx} 
\usepackage{fancyhdr}
\usepackage{hyperref}
\usepackage{amssymb,amsmath,amsfonts}
\usepackage[center]{caption}
\usepackage{suffix}
\usepackage{imakeidx}
\usepackage{balance}
\usepackage{epstopdf}
\usepackage{color}

\ifCLASSINFOpdf
\else
\fi
\begin{document}
%
\title{Power Allocation and User Assignment Scheme for Beyond 5G Heterogeneous Networks}
%
%
%

\author{Khush Bakht,
        Furqan Jameel,
        Zain Ali,
				Wali Ullah Khan,
				Imran Khan,
				Guftaar Ahmad Sardar Sidhu, and
				Jeong Woo Lee
\thanks{Khush Bakht is with the Department of Electronic Engineering, Fatima Jinnah Women University Rawalpindi 46000, Pakistan (email: k.bakht1990@gmail.com).}
\thanks{Furqan Jameel is with the Department of Communications and Networking, Aalto University, 02150 Espoo, Finland. (email: furqanjameel01@gmail.com).}
\thanks{Zain Ali, Guftaar Ahmad Sardar Sidhu are with the Department of Electrical Engineering, COMSATS University, Islamabad 45550, Pakistan (email: zainalihanan1@gmail.com, guftaarahmad@comsats.edu.pk).}
\thanks{Wali Ullah Khan is with the School of Information Science and Engineering, Shandong University, Qingdao 266237, China (email: waliullahkhan30@gmail.com).}
\thanks{Imran Khan is with the University of Engineering and Technology, Peshawar Pakistan (email: ikeu007@gmail.com).}
\thanks{Jeong Woo Lee is with the School of Electrical and Electronics Engineering, Chung-Ang University, Seoul, South Korea (email: jwlee2@cau.ac.kr).}
\thanks{}}

%
%

\markboth{}%
{\MakeLowercase{\textit{et al.}}}
%



\maketitle

\begin{abstract}
The issue of spectrum scarcity in wireless networks is becoming prominent and critical with each passing year. Although several promising solutions have been proposed to provide a solution to spectrum scarcity, most of them have many associated tradeoffs. In this context, one of the emerging ideas relates to the utilization of cognitive radios (CR) for future heterogeneous networks (HetNets). This paper provides a marriage of two promising candidates (i.e., CR and HetNets) for beyond fifth generation (5G) wireless networks. More specifically, a joint power allocation and user assignment solution for the multi-user underlay CR-based HetNets has been proposed and evaluated. To counter the limiting factors in these networks, the individual power of transmitting nodes and interference temperature protection constraints of the primary networks have been considered. An efficient solution is designed from the dual decomposition approach, where the optimal user assignment is obtained for the optimized power allocation at each node. The simulation results validate the superiority of the proposed optimization scheme against conventional baseline techniques.
\end{abstract}
%
\begin{IEEEkeywords}
Beyond 5G, Cognitive Radio (CR), Dual Decomposition, User Fairness, Heterogeneous Networks (HetNets)
\end{IEEEkeywords}
%
%
\IEEEpeerreviewmaketitle
\section{Introduction}
In the last few years, the wireless systems have evolved to the point where a homogeneous cellular networks have achieved near optimal performance \cite{8340813}. These advancements in the homogeneous cellular networks, though significant, may not be enough to support beyond fifth generation (5G) wireless networks \cite{Gao}. To do so, dynamic and exhaustive improvements in spectral efficiency are needed. \textcolor{black}{One of the proposed solutions is advanced networks which consist of multiple tiers of high-powered base stations (BSs) overlaid with low-powered BS both having different coverage areas, also known as heterogeneous networks (HetNets) \cite{1het}.} Typically, a HetNet consists of main macro base stations (MBSs), a few pico base stations (PBSs), and several femto base stations (FBSs). An MBS in HetNets has a high transmission power and greater coverage area which is then overlaid with low-powered PBS and FBS \cite{xu2018robust,2a}. 

There are different purposes of PBS and FBS in the HetNets. \textcolor{black}{Generally speaking, the PBS are laid in a  dense traffic areas to improve coverage in hotspot areas, while FBS are overlaid in a manner to remove the coverage issues in homogeneous systems, thereby improving the overall performance.} In conventional homogeneous cellular network, the mobile terminal is associated to a BS on the basis of downlink signal-to-interference-and-noise ratio (SINR). \textcolor{black}{Specifically, a user is associated with a particular BS which offers SINR greater than the other BSs \cite{1}.} In HetNets, the SINR-based association principle leads to having load balancing issue among MBS and PBS. \textcolor{black}{Hence, the network resources need to be intelligently distributed among BSs for higher spectral efficiency \cite{2a,1}.} 

\subsection{Related Works}

By the end of 2020, it is anticipated that up to 50 billion devices will exist in the world including static and mobile platforms \cite{8581856}. Due to this reason, there has been an upsurge in the research on HetNets to provide efficient and long-term solutions \cite{sun2018energy}. Zhang \emph{et al.} in \cite{3} investigated the problem of user association in HetNets. They considered an optimization problem with different traffic capacity limits, quality of service (QoS) requirements and power budget constraints. The proposed scheme of user association improved the performance of the network. \textcolor{black}{A bit-rate adoption method was proposed by the authors of \cite{cwalina2018novel} for body area HetNets. Their scheme showed the promise of increasing the efficiency of packet transmission regardless of placement of the node.} The authors of \cite{4} performed joint optimization of resource allocation and user association in HetNets. They presented and compared three allocation strategies, i.e., orthogonal deployment, co-channel deployment, and partially shared deployment. \textcolor{black}{Sequential quadratic programming (SQP)-based power optimization approach was presented in \cite{Khan} to maximize the sum rate of small cell networks.} \textcolor{black}{Similar works have been done by the authors of \cite{7}, where a pricing-based approach for associating the user to a particular BS was proposed by the authors of \cite{8} and the idea of a distributed price update strategy was presented for achieving the user fairness.} 
 
Liu \emph{et al.} presented a fair user association scheme in \cite{9}. \textcolor{black}{They used the Nash bargaining solution (NBS), whereby, the optimization with fairness was achieved among competing BSs.} In \cite{11}, the HetNets using orthogonal frequency division multiple access (OFDMA) network were presented by the authors. The aim was to manage radio resource by maximizing the throughput of user having minimum rate. \textcolor{black}{From the perspective of non-orthogonal multiple access (NOMA), the author of \cite{khan2019efficient} provided a novel idea to use interference-aided vehicular networks and highlighted some key challenges.} Similarly, the authors of \cite{12} managed the radio resources by a scheduling algorithm implemented by a central global resource controller (GRC). The algorithm optimized the attributes such as fairness among users, spectral efficiency and battery lifetime. Relay-based HetNets were considered by the authors of \cite{qin2019cross}, wherein, they proposed a method to suppress inter-cell and intra-cell interference. Their proposed scheme was shown to outperform existing baseline methods in terms of sum-rate performance. \textcolor{black}{Another similar and much recent work \cite{jabeen2019joint} provided optimal time switching and power splitting technique for improving the performance of wireless network.} \textcolor{black}{The authors in \cite{khodmi2019joint} jointly optimized power allocation and user association in ultra dense heterogeneous networks using non-cooperative game theory. Thus achieving the increase in system throughput as well as optimal power allocation.}

Semov \emph{et al.} proved that in HetNets by taking the geographical position into account the users throughput and fairness can be improved. \cite{13}. A similar concept was employed in the form of relays by the authors of \cite{14} to ensure proportional fairness among user equipment by taking into account backhaul links between the relays and BS. \textcolor{black}{On the other hand, the authors of \cite{20} considered distributed antenna system (DAS) and compared co-channel resource allocation schemes for energy-efficient communication}. In \cite{9f}, the authors studied a single cell HetNet having one macro and one picocell for efficient resource allocation such that the energy efficiency is maximized by proposing an iterative resource allocation algorithm. \textcolor{black}{However, the same authors did not consider multi-cellular network for the evaluation.} To achieve the spectral efficiency of the system more advanced dynamic spectrum access techniques (DSA) should be employed. The cognitive radio (CR) is an efficient DSA technique\cite{15} that allows secondary (unlicensed) users (SU) to access the spectrum of primary (licensed) users (PU) in an opportunistic way\cite{16}. In CR, spectrum sharing can be classified as spectrum overlay and spectrum underlay. \textcolor{black}{In spectrum overlay, the SU can transmit simultaneously on the frequency band used by the PU through adjusting its transmit power such that it does not cause much interference for the PU \cite{16}.} 
 
Of late, the CR-based HetNets have been drawing a lot of research attention nowadays due to their dynamic resource allocation property. The power adjustment of BS and mobile users can be achieved by dynamic resource allocation \cite{7f}. In this regard, the author in \cite{18} maximized the energy efficiency subject to power and interference constraint in OFDMA based CR networks. The authors applied the convex optimization theory and proposed an iterative algorithm. \textcolor{black}{In \cite{8f}, the authors considered a cognitive femtocell network using OFDMA and solved a sum-utility maximization and dynamic resource allocation problem with the help of dual decomposition method.} \textcolor{black}{In \cite{liu2018energy} the authors considered the stochastic optimization model for maximizing the long term energy efficiency in time varying heterogeneous networks. The proposed problem is a mixed integer problem and the Lagrange dual method has been used to solve it.} In \cite{14f}, the authors considered the resource allocation problem for rate maximization in multi-user cognitive heterogeneous networks. The authors considered the maximum transmit power of cognitive microcell base station and cross-tier interference constraints, simultaneously. They converted the non-convex optimization problem into a geometric programming problem and solved it in a distributed way using the Lagrange dual method.

\begin{table}
\centering
\caption{List of acronyms.}
\label{tab1}
\begin{tabular}{|c|c|} \hline 
\textbf{Acronym} & \textbf{Definition}\\
\hline
5G & Fifth Generation \\ \hline
AWGN &  Additive White Gaussian Noise  \\ \hline
BS & Base Station\\ \hline
CR & Cognitive Radio \\ \hline
D2D & Device to Device\\ \hline
DAS & Distributed Antenna System\\ \hline
DSA & Dynamic Spectrum Access \\ \hline
FBS & Femto Base Station\\ \hline
GRC & Global Resource Controller\\ \hline
HetNet & Heterogeneous Networks\\ \hline
KKT & Karush Kuhn Tucker\\ \hline
MBS & Macro Base Station\\ \hline
NBS & Nash Bargaining Solution  \\ \hline
OFDMA & Orthogonal Frequency Division Multiple Access \\ \hline
PBS & Pico Base Station \\ \hline
PSD & Power Spectral Density\\ \hline
PU & Primary User \\ \hline
QoS & Quality of Service \\ \hline
RRM & Radio Resource Management\\ \hline
SQP & Sequential Quadratic Programming \\ \hline
SU &  Secondary User \\ \hline
SINR & Signal to Interference Plus Noise Ratio  \\ \hline
\end{tabular}
\end{table}

\subsection{Motivation and Contributions}

Although the research works reported in recent years have considered the overall system's performance maximization for CR-based HetNets, the problem of fairness among different users has received little attention. The potential problem arises when the schemes proposed for the sum-rate maximization assign very few or no resources to some of the users with higher fading conditions. This uneven distribution of resources results in degrading the achievable performance for different users. This problem may become more serious for the CR-based HetNets due to the limiting factor of increased interference. Thus, optimization of the transmission for user fairness under more practical constraints becomes essential. To the best of our knowledge, the resource optimization and user assignment techniques for fair rate allocation have not been jointly investigated in the literature due to the higher level of complexity involved in finding the optimal solution.  

To fill this gap in the literature and provide a comprehensive solution to the user fairness problem in CR-based HetNets, we provide a joint strategy for power allocation and user fairness. In particular, we consider the joint power optimization and user association problem in HetNets for achieving fairness among different users. We first formulate a joint optimization problem subject to power and interference constraints. Then, to provide a less complex and an efficient solution, we design an algorithm from the dual decomposition strategy. The presented numerical results indicate the importance and utility of our scheme in comparison to the baseline methodologies.

\subsection{Organization}

The remainder of this paper is organized as follows. The system model and problem formulation are presented in Section 2. In Section 3, the proposed solution has been described, while Section 4 discusses the simulation results. Finally, Section 5 presents some concluding remarks and future research directions. In addition, the list of acronyms used throughout this paper has been provided in Table \ref{tab1}.   
      
\begin{figure*}
  \centering
      \includegraphics[width=0.9\textwidth]{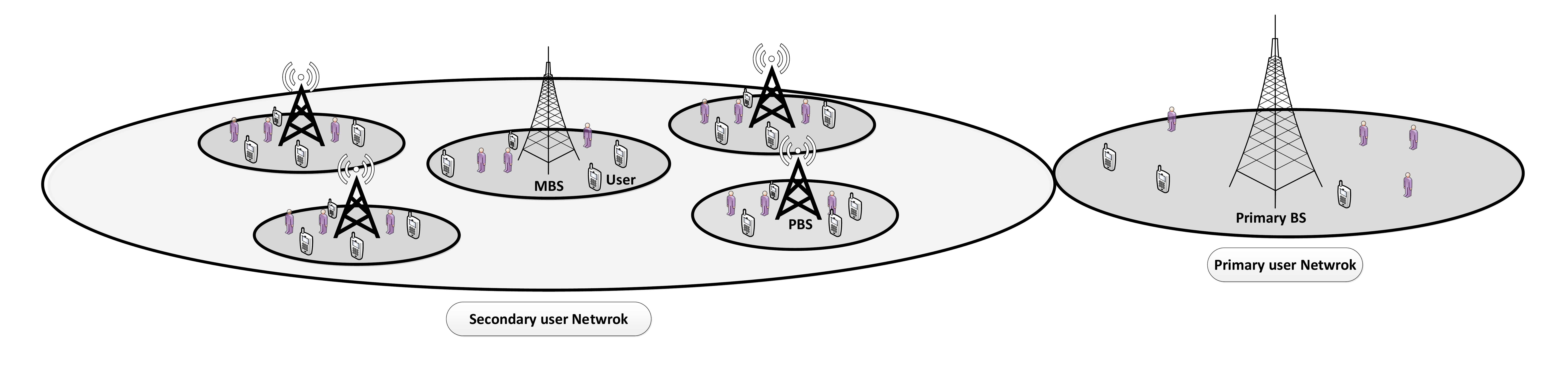}
      \caption{Heterogeneous Network Using Cognitive Radio Spectrum in Underlay Mode}
      \label{Fig.1}
      \end{figure*} 

\section{System Model and Problem Formulation}
This section describes the considered system model and provides the steps related to problem formulation.
\subsection{System Model}
A downlink CR transmission is considered, where the secondary HetNet system is reusing the spectrum of the primary network in an underlay mode, as shown in Fig. \ref{Fig.1}. The HetNet consists of a single macrocell overlaid with multiple pico BSs, intended to transmit data to multiple users. \textcolor{black}{It is assumed that all the devices are equipped with a single antenna and experience independent and identically distributed (i.i.d) Rayleigh fading.} The channel bandwidth is distributed among MBS and PBSs in such a way that MBS and PBSs are non-orthogonal to each other, while each PBS is orthogonal to other PBSs in the HetNet.

In this model, BSs are denoted as $BS_b$, such that $\mathbb{B}={BS_b|b=1,2,3,...,B}$, where $BS_1$ is modeled as macro and the remaining BSs act as pico BSs. Further, there are $A$ users and $C$ channels allocated to BS. The signal-to-interference-and-noise ratio (SINR) of $a$-th user from the $b$-th BS at the $c$-th channel is given as in \cite{4}
\begin{align}
SINR_{a,c,b} = \frac{ P_{a,c,b} g_{a,c,b}}{P'_{a,c,b'}  f_{a,c,b}+\sigma_a^2}, 
\end{align}
where $P_{a,c,b}$ is the transmit power for $a$-th user connected to $b$-th BS at $c$-th channel, $\alpha_{a,c,b}$ denotes the association variable which belongs to set ${ \{0,1} \}$, $\sigma_a^2$ is the power spectral density (PSD) of the noise while $g_{a,c,b}$ and $f_{a,c,b'}$ are the channel gains from intended BS to user and from interfering BS to the user, respectively. Furthermore, $P'_{a,c,b'}$ denotes the transmit power of interfering BS. 

\subsection{Problem Formulation}               
One of our key objectives in this article is fair maximization of the data rate of each user in the network. To facilitate mathematical analysis, we introduce a binary variable $\alpha_{a,c,b}$, such that
\begin{align}
\alpha_{a,c,b}=\!\begin{cases}
               1,\,\,\, &\text{when $a$-th user is associated with $b$-th BS}\\ &\text{through $c$-th channel,}\\
              0,\,\,&\text{otherwise,} \nonumber
            \end{cases}.
\end{align}

Based on the above expression, the channel allocation follows 
\begin{align}
&\sum\limits_{c=1}^C \sum\limits_{b=1}^B \alpha_{a,c,b}=1\,\,\,\quad\quad\quad \, \forall\ a={1,2,3,...A}.\label{a3}
\end{align}

To protect PUs from the interference and to improve the performance of the system, resource allocation at BSs need to be performed such that
\begin{align}
\sum\limits_{a=1}^A \sum\limits_{c=1}^C \sum\limits_{b=1}^B \alpha_{a,c,b} P_{a,c,b} h_{a,c,b}\leq I_{th},\label{a4}
\end{align}
where $P_{a,c,b}$ is the power allocated by $b$-th BS to $c$-th channel allocated to $a$-th user and $h_{a,c,b}$ represents the gain of interference channel $c$ from $b$-th BS to primary BS. Finally, to ensure that the power consumed by each BS is less than or equal to the power budget, power allocation at each BS needs to be ensured as
\begin{align}
\sum\limits_{a=1}^A \sum\limits_{c=1}^C P_{a,c,b}\alpha_{a,c,b}\leq P_b\quad\ \forall\ b={1,2,3,...B}.\label{a5}
\end{align}

To achieve fairness in rates of different users, a max-min-based optimization framework is adopted such that the problem is stated, mathematically as
\begin{align}
\textbf{P1:}\quad&\max_{P_{a,c,b}, \alpha_{a,c,b}} \min\,\, \log_2 \left(1+SINR_{a,c,b} \right),\\        
&\text{s.t.}\quad (\ref{a3}), (\ref{a4})\,\, \text{and}\,\,(\ref{a5})\nonumber,
\end{align}
where $P_b$ denote total power available at $b$-th BS and $I_{th}$ represents the sum interference threshold. Here, the first constraint ensures that the total power allocated by $b$-th BS must be within the available power budget. \textcolor{black}{Similarly, the second constraint esures that the primary system is well protected from interference due to the underlay communication, and the third constraint makes sure that $c$-th channel is allocated to just one SU.}

\section{Proposed Solution}

To provide a viable solution to the aforementioned problem, we propose a solution based on duality theory. As is evident from the above expression, the allocation of channels and power loading are strongly coupled variables, thus, a joint optimization approach is needed. The problem \textbf{P1} is a mixed binary integer programming and requires an exhaustive search. Fortunately, it has been shown by \cite{zain} that for a sufficiently large number of subchannels the gap between the dual solution and the primal solution reduces to zero, regardless of the non-convexity of the original problem. Thus, we exploit the duality theory to decompose and optimally solve the coupled problem.
 
To convert the complex max min problem into standard optimization form, we introduce intermediate variable $t$ such that 
\begin{align}
t\ \leq \log_2 \left(1+SINR_{a,c,b} \right)\quad \forall a,c,b,\label{a10}
\end{align}

The introduction of the intermediate variable transforms the problem as
\begin{align}
\textbf{P2:}\quad&\max_{P_{a,c,b},t} \ t  \label{obj1}\\
&\text{s.t.} \nonumber\quad  (\ref{a3}), (\ref{a4}), (\ref{a5})\,\,\text{and}\,\,(\ref{a10}).
\end{align}

To obtain an immediate solution of auxiliary variables, from the structure of objective in (\ref{obj1}), we utilize the fact that for any $y \geq 0$, minimizing $y$ is equivalent to minimization of $y^{2}$. It is a known fact that maximizing $y$ is equal to minimizing $-y$. Hence, to transform the problem into a standard minimization problem we replace the objective in (\ref{obj1}) with its negative. \textcolor{black}{After making these transformations \textbf{P2} is written as}
\begin{align}
\textbf{P3:}\quad&\min_{P_{a,c,b},t} -t^{2}  \label{obj2}\\
&\text{s.t.} \nonumber\quad  (\ref{a3}), (\ref{a4}), (\ref{a5})\,\,\text{and}\,\,(\ref{a10}).
\end{align}

Now, substituting $x=-t$, we obtain
\textcolor{black}{\begin{align}
&\min_{P_{a,c,b},t} x^{2}  \label{obj3}\\
&\text{s.t.} \nonumber\quad  (\ref{a3}), (\ref{a4}), (\ref{a5})\,\, \\
&\text{and}\,\, -x\ \leq \log_2 \left(1+SINR_{a,c,b} \right)\, \forall a,c,b. \nonumber
\end{align}}
 
The dual function associated with (\ref{obj3}) is given by
\begin{align}
&D \left(\lambda_{a}, \eta_{b},v \right)= \min_{x,P_{a,c,b},\alpha_{a,c,b}}\nonumber\\&\sum\limits_{a=1}^A\lambda_{a}\Big(-x- \sum\limits_{c=1}^C \sum\limits_{b=1}^B\alpha_{a,c,b} \log_2\! \left(1+SINR_{a,c,b}\right)\Big) \nonumber\\&+  x^{2} +\sum\limits_{b=1}^B \eta_b \Bigg(\sum\limits_{a=1}^A \sum\limits_{c=1}^C\alpha_{a,c,b} P{a,c,b}-P_{b} \Bigg) \nonumber\\& +v \left(\sum\limits_{a=1}^A \sum\limits_{c=1}^C \sum\limits_{b=1}^B \alpha_{a,c,b}P_{a,c,b}h_{a,c,b} -I_{th}\right), \label{aa1}\\
&\text{s.t.}\quad\sum\limits_{c=1}^C \sum\limits_{b=1}^B \alpha_{a,c,b}=1,\ \forall\ a={1,2,3,...A}.\nonumber
\end{align}

The expression in (\ref{aa1}) can be written as
\begin{align}
&D \left(\lambda_{a}, \eta_{b},v \right)= \min_{x,P_{a,c,b},\alpha_{a,c,b}}x^{2} +\sum\limits_{a=1}^A\sum\limits_{c=1}^C \sum\limits_{b=1}^B\alpha_{a,c,b}\nonumber\\&\bigg(-\lambda_{a}\log_2  \left(1\!\!+\!SINR_{a,c,b} \right) + \eta_b P{a,c,b} +v P_{a,c,b}h_{a,c,b}\bigg) \nonumber\\&\!\! -\!\sum\limits_{a=1}^A\!\lambda_{a} x - \sum\limits_{b=1}^B\eta_{b} P_{b}-v I_{th}\label{aa2},\\
&\text{s.t.}\quad\sum\limits_{c=1}^C \sum\limits_{b=1}^B \alpha_{a,c,b}=1,\ \forall a={1,2,3,...A}.\nonumber
\end{align}

For any given channel allocation, dual decomposition guides to solve the following sub-problems
\begin{align}
\textbf{P4:}\quad&\min_{x} \ \bigg(x^2- \ x   \sum\limits_{a=1}^A \lambda_{a} \bigg), \label{22}
\end{align}
\begin{align}
\textbf{P5:}\quad&\min_{P_{a,c,b}} \Big(-\lambda_{a} \log_2 (1+SINR_{a,c,b} \nonumber\\&+ \eta_b P_{a,c,b} + v P_{a,c,b} h_{a,c,b} \Big). \label{23a}
\end{align}

Using the KKT conditions to find the solution of the problems given in (\ref{22}) and (\ref{23a}), we get
\begin{align}
x^{*}=  \frac{1}{2} \bigg( \sum\limits_{a=1}^A \lambda_{a} \bigg)^{+},
\end{align}
\begin{align}
P_{a,c,b}^{*}= \left(\frac{\Phi_{a,c,b}  -  P_{a,c,b}' \alpha_{a,c,b} f_{a,c,b} }{g_{a,c,b} \left( \eta_b + v h_{a,c,b} \right)}\right)^{+}, 
\end{align} 
where $\Phi_{a,c,b}=\lambda_{a} g_{a,c,b} + \sigma_a^2 \eta_b + v h_{a,c,b}$, for all $a,c,b$, when $(\Psi)^{+}=\max(0,\Psi)$. The detailed derivation steps have been provided in the Appendix. 

Now, to find the optimum value of $\alpha_{a,c,b}$, following optimization problem is considered
\begin{align}
&\min_{\alpha_{a,c,b}}\sum\limits_{a=1}^A \sum\limits_{c=1}^C \sum\limits_{b=1}^B \alpha_{a,c,b} \bigg(-\lambda_{a} \log_2 \Big(1+SINR_{a,c,b} \Big)\nonumber\\& +  \eta_b P_{a,c,b}^{*}+ v P_{a,c,b}^{*} h_{a,c,b} \bigg), \label{23}\\
&\text{s.t.}\quad\sum\limits_{c=1}^C \sum\limits_{b=1}^B \alpha_{a,c,b}=1,\ \forall a={1,2,3,...A}.\nonumber 
\end{align}

Evidently, the optimal solution can be found, such that
\begin{align}
\alpha_{a,c,b}^{*}=\begin{cases}
              & \text{for}\, a=\text{arg} \min_{c}\!\Big(-\lambda_{a} \log_2 (1+SINR_{a,c,b})\nonumber\\& +  \eta_b P_{a,c,b}^{*}+ v P_{a,c,b}^{*} h_{a,c,b} \Big),\\
              &\text{otherwise}. 
            \end{cases}.\tag{18}
\end{align}

The dual problem is convex, hence, sub-gradient method can be adopted to find the solution. The dual variables are updated at each iteration as
\begin{align}
&\lambda_{a}^{ \left(itr \right)}=\bigg(\lambda_{a}^{ \left(itr-1 \right)}+\delta^{ \left(itr-1 \right)} \nonumber \\
&\times \Big(-\sum\limits_{b=1}^B \sum\limits_{c=1}^C\log_{2}(1\!\!+\!SINR_{a,c,b})-x \Big) \bigg),\tag{19}\\
&\eta_b^{ \left(itr \right)}=  \bigg(\eta_b^{ \left(itr-1 \right)}+ \delta^{ \left(itr-1 \right)} \Big( \sum\limits_{a=1}^A \sum\limits_{c=1}^C  P_{a,c,b}-P_b \Big) \bigg),\tag{20}\\
& v^{ \left(itr \right)}=\bigg(v^{ \left(itr-1 \right)}+\delta^{ \left(itr-1 \right)} \nonumber \\
&\times \Big(\sum\limits_{a=1}^A \sum\limits_{c=1}^C \sum\limits_{b=1}^B P_{a,c,b} h_{a,c,b}-I_{th}\Big) \bigg), \tag{21}
\end{align}
where $\delta$ is the step size. To obtain the joint optimization solution, power loading and channel allocation are updated in each iteration.
\section{Simulation Results}
In this section, we present the performance of the proposed scheme. The simulation environment for the proposed schemes is MATLAB. The noise PSD is $\sigma_a^2= 0.1$. For simulation we have considered the maximum number of users in the system is $A=30$, the total channels available in the system is $C=20$ and the total BS available in the system is $B=5$. For the sake of the evaluation, we have compared our proposed optimal allocation and optimal power scheme (OAOP) with baseline schemes, i.e., fixed channel allocation and optimal power loading (FAOP) and fixed channel allocation and fixed power loading (FAFP).
    
\begin{figure}[!t]
\centering
\includegraphics [scale=.4]{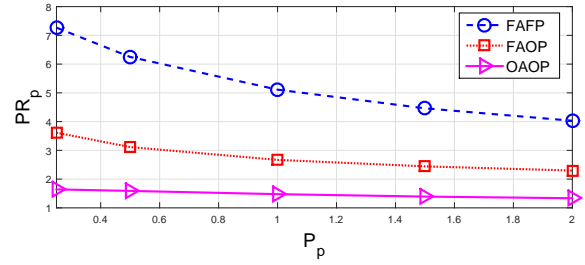}
\centering \caption{Peak to average ratio with changing picocell power.}
\label{Fig.2}
\end{figure}

In Fig. \ref{Fig.2}, peak to average rate ratio (PR) has been plotted against different parameters. Different values of PR show fairness among the users. \textcolor{black}{The smaller values of PR indicates a more fair scheme.} Moreover, the effect of changing picocell power $PR_p$ on PR has also been shown. \textcolor{black}{The picocell power has been varied from 0.25 to 2.} Considering $P_p=0.25$, the percentage gap between FAFP and FAOP is $49.74\%$, between FAFP and OAOP is $22.59\%$ and between FAOP and OAOP is $45.42\%$. \textcolor{black}{It can be seen that the value of PR decreases with an increase in the power. This indicates that high picocell power causes high interference which results in decreasing the PR. Moreover, a comparison of the three schemes clearly indicates that the OAOP scheme is the fairest among all.} This is because optimization on channel allocation along with optimal power loading makes the problem more flexible compared to the cases when only power loading is optimized and if channel allocation and power loading are fixed. Also for $P_p=0.5$ the percentage gap between FAFP and FAOP is $49.83\%$, between FAFP and OAOP is $25.44\%$ and between FAOP and OAOP is $51.05\%$. Finally, increasing the value to $P_p=2$, the percentage gap between FAFP and FAOP is $56.99\%$, between FAFP and OAOP is $33.12\%$ and between FAOP and OAOP is $58.11\%$.  

\begin{figure}[!t]
\centering
\includegraphics [scale=.4]{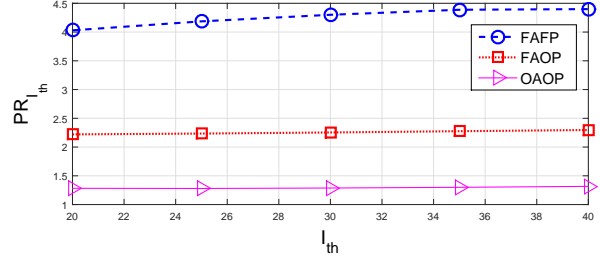}
\caption{Peak to average ratio with changing $I_{th}$.}
\label{Fig.3}
\end{figure}

Fig. \ref{Fig.3} is showing the PR with changing interference constraint $PR_{I_{th}}$. The interference constraint has been varied from 20 to 40. Considering $I_{th}=20$, the percentage gap between FAFP and FAOP is $55.08\%$, between FAFP and OAOP is $31.83\%$ and between FAOP and OAOP is $57.79\%$. Also for $I_{th}=25$ the percentage gap between FAFP and FAOP is $53.39\%$, between FAFP and OAOP is $30.57\%$ and between FAOP and OAOP is $57.27\%$. Moreover, if $I_{th}=30$, then the percentage gap between FAFP and FAOP is $52.41\%$, between FAFP and OAOP is $29.95\%$ and between FAOP and OAOP is $57.14\%$. Further increasing the value of $I_{th}=35$, the percentage gap between FAFP and FAOP is $52.00\%$, between FAFP and OAOP is $29.62\%$ and between FAOP and OAOP is $57.11\%$. Finally, increasing the value to $I_{th}=40$, the percentage gap between FAFP and FAOP is $51.86\%$, between FAFP and OAOP is $28.91\%$ and between FAOP and OAOP is $56.31\%$. Comparison of schemes in terms of the percentage gap shows that as the interference constraint $I_{th}$ increases the percentage gap is decreasing. Moreover, the average gap between the graphs of FAFP with FAOP and FAOP with OAOP scheme is near to each other. Graphs also show that OAOP outperforms all other schemes in terms of fairness. The reason behind this is the additional flexibility provided by optimum channel allocation. 

\begin{figure}[!t]
\centering
\includegraphics [scale=.4]{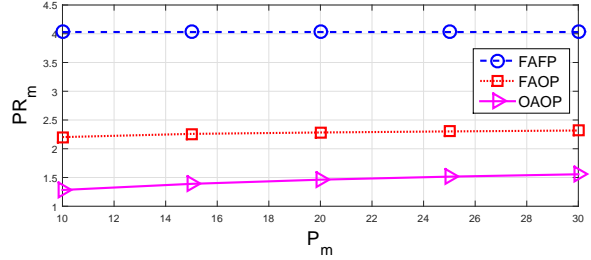}
\caption{Peak to average ratio with changing macro power.}
\label{Fig.4}
\end{figure}

The effect of changing the power of macro base station $PR_m$ on the peak to average rate ratio is shown in Fig. \ref{Fig.4}. The macro power has been varied from 10 to 30. Considering $PR_m=10$, the percentage gap between FAFP and FAOP is $54.64\%$, between FAFP and OAOP is $31.83\%$ and between FAOP and OAOP is $58.26\%$. Also for $PR_m=15$ the percentage gap between FAFP and FAOP is $56.05\%$, between FAFP and OAOP is $34.46\%$ and between FAOP and OAOP is $61.48\%$. Moreover, if $PR_m=20$, then the percentage gap between FAFP and FAOP is $56.65\%$, between FAFP and OAOP is $36.27\%$ and between FAOP and OAOP is $64.03\%$. Further increasing the value of $PR_m=25$, the percentage gap between FAFP and FAOP is $57.09\%$, between FAFP and OAOP is $37.59\%$ and between FAOP and OAOP is $65.84\%$. Finally, increasing the value to $PR_m=30$, the percentage gap between FAFP and FAOP is $57.49\%$, between FAFP and OAOP is $38.61\%$ and between FAOP and OAOP is $67.15\%$. Hence, comparing all the three schemes show that the OAOP is the superior scheme. 

\begin{figure}[!t]
\centering
\includegraphics [scale=.4]{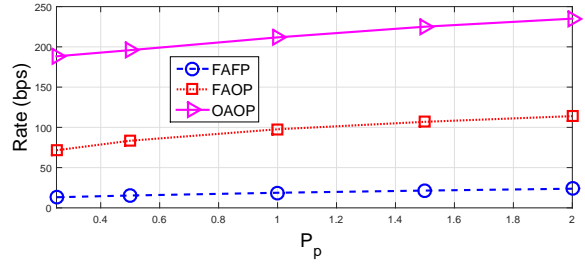}
\caption{Sum throughput with changing pico power.}
\label{Fig.5}
\end{figure}

In the results presented here, the sum throughput has also been maximized along with achieving fairness among the users. Fig. \ref{Fig.5} shows the sum throughput versus pico power $P_p$. The $P_p$ has been varied from 0.25 to 2. The percentage gap of OAOP scheme with FAOP is $38\%$, between OAOP with FAFP is $7\%$ and between FAOP with FAFP is $18.42\%$ where the value of $P_p=0.25$. \textcolor{black}{The results evidently show that the spectral efficiency of OAOP schemes is the most optimal among the other power allocation schemes. This is because when channel allocation is optimized and the channels are allocated to users in such a way that would maximize the total data rate of users.} Then, increasing the value of $P_p=0.5$ percentage gap of OAOP with FAOP is $42.56\%$, between OAOP with FAFP is $7.825\%$ and between FAOP with FAFP is $18.38\%$. Further increasing the value to $P_p=2$ OAOP with FAOP is $48.51\%$, between OAOP with FAOP is $10.12\%$ and between FAOP with FAFP is $20.85\%$.

\begin{figure}[!t]
\centering
\includegraphics [scale=.4]{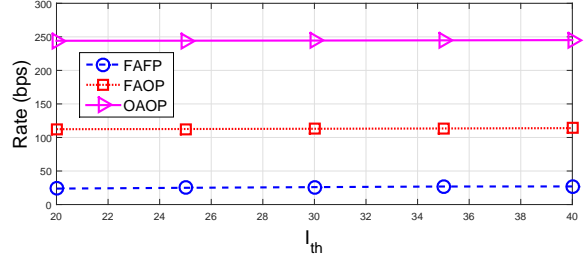}
\caption{Sum throughput with changing $I_{th}$.}
\label{Fig.6}
\end{figure}

Fig. \ref{Fig.6} shows the sum throughput versus $I_{th}$. The percentage gap of OAOP scheme with FAOP is $45.98\%$, between OAOP with FAFP is $9.73\%$ and between FAOP with FAFP is $21.17\%$ where the value of $I_{th}=20$. Further increasing the value of $I_{th}=25$ percentage gap of OAOP with FAOP is $46.09\%$, between OAOP with FAFP is $10.22\%$ and between FAOP with FAFP is $22.17\%$. Moreover, if $I_{th}=30$, percentage gap between OAOP with FAOP is $46.21\%$, between OAOP with FAFP is $10.62\%$ and between FAOP with FAFP is $23\%$. Considering $I_{th}=35$, OAOP with FAOP is $46.32\%$, between OAOP with FAFP is $10.98\%$ and between FAOP with FAFP is $23.71\%$. Further increasing the value to $I_{th}=40$ OAOP with FAOP is $46.47\%$, between OAOP with FAOP is $11.01\%$ and between FAOP with FAFP is $23.71\%$. Hence, the sum throughput of OAOP scheme is maximum.

\begin{figure}[!t]
\centering
\includegraphics [scale=.4]{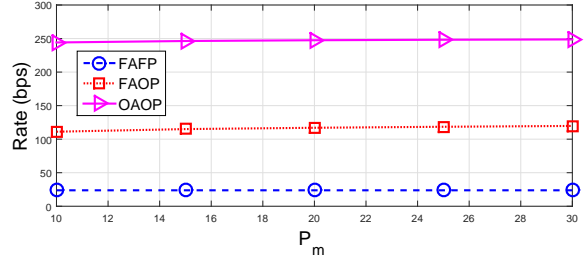}
\caption{Sum throughput with changing macro power.}
\label{Fig.7}
\end{figure}

The effect of varying macro power $P_m$ on the sum throughput is shown in Fig. \ref{Fig.7}. The $P_m$ has been varied from 10 to 30. The percentage gap of OAOP scheme with FAOP is $45.49\%$, between OAOP with FAFP is $9.73\%$ and between FAOP with FAFP is $21.40\%$ where the value of $P_m=10$. Further increasing the value of $P_m=15$ percentage gap of OAOP with FAOP is $46.69\%$, between OAOP with FAFP is $9.65\%$ and between FAOP with FAFP is $20.67\%$. Moreover, if $P_m=20$, percentage gap between OAOP with FAOP is $47.27\%$, between OAOP with FAFP is $9.61\%$ and between FAOP with FAFP is $20.32\%$. Considering $P_m=25$, OAOP with FAOP is $47.74\%$, between OAOP with FAFP is $9.58\%$ and between FAOP with FAFP is $20.06\%$. Further increasing the value to $P_m=30$ OAOP with FAOP is $48.13\%$, between OAOP with FAOP is $9.56\%$ and between FAOP with FAFP is $19.86\%$. The results show that OAOP scheme is the most optimal.

\begin{figure}[!t]
\centering
\includegraphics [scale=.4]{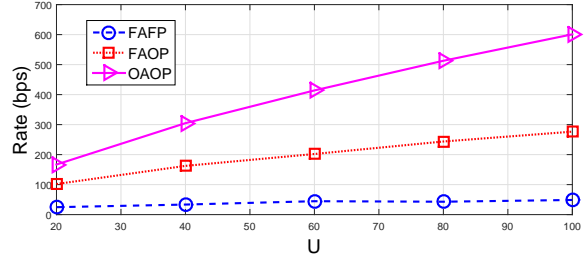}
\caption{Sum throughput with increasing users}
\label{Fig.8}
\end{figure}

Fig. \ref{Fig.8} shows the values of sum throughput with a changing number of users $U$. \textcolor{black}{From this plot, we identified that the performance of OAOP is far superior to other schemes which is due to the fact that adding new users to the system introduces new channel gains. This results in increasing the flexibility of the problem. In OAOP, both channel allocation and power loading are optimized, therefore, this scheme is better at taking advantage of the newly added users in the network. In FAOP only power loading is optimized so the addition of new users increases the data rate because of the reduced degree of freedom compared to OAOP. A very slight increase is observed in FAFP when new users are added, as, neither channel allocation nor power loading is optimized in this scheme.} The $U$ has been varied from 20 to 100. Considering $U=20$, the percentage gap of OAOP scheme with FAOP is $61.07\%$, between OAOP with FAFP is $14.82\%$ and between FAOP with FAFP is $24.27\%$. Further increasing $U=40$ percentage gap of OAOP with FAOP is $53.28\%$, between OAOP with FAFP is $10.99\%$ and between FAOP with FAFP is $20.63\%$. Moreover, if $U=60$, the percentage gap between OAOP with FAOP is $48.81\%$, between OAOP with FAFP is $10.79\%$ and between FAOP with FAFP is $20.10\%$.  

\begin{figure}[!t]
\centering
\includegraphics [scale=.33]{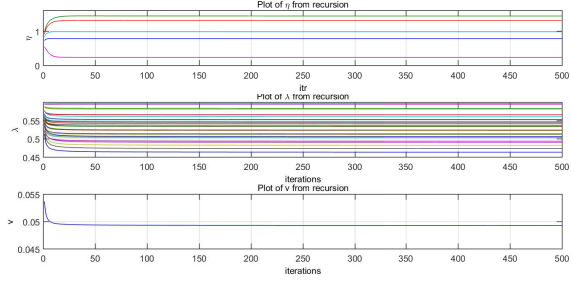}
\caption{Convergence of $\eta,\lambda,v$ with increasing iterations.}
\label{Fig.9}
\end{figure}

Fig. \ref{Fig.9} shows the convergence of dual variable $\lambda_a$, $\eta$ and $v$ with initial value as 0.6, $\eta$ as 0.6 and v as 0.1. The smaller the value of step size more fine the convergence is, a bigger value leads to fast convergence.
\section{Conclusion}
This paper optimized the power allocation and user association to achieving fairness among different users in underlay CR-based HetNets. Specifically, we adopted a max-min-based fairness framework under various practical constraints. Problem was first transformed into a standard maximization optimization and then dual decomposition is used to solve the integer programming problem. The dual problem was solved through the sub-gradient method. The power allocation at each node is obtained from the convex optimization techniques for the known power allocation at all other nodes. The results are compared with the sub-optimal scenarios when the fixed channel allocation is assumed for the optimized/non-optimized power allocation. The results showed that the proposed scheme outperforms all other candidates and the spectral efficiency has been improved due to the simultaneous transmission with the primary network.

\section*{Appendix}

The Lagrangian L associated with the optimization problem \textbf{P5} is
\begin{align}
&\min_{P_{a,c,b}} \left(-\lambda_{a} \log_2 \left(1+SINR_{a,c,b} \right) + \eta_b P_{a,c,b} + v P_{a,c,b} h_{a,c,b} \right). \nonumber
\end{align}

In the above expression, we have converted our problem from a constrained problem to an unconstrained problem.
\begin{align}
&
\min_{P_{a,c,b \geq 0}} \Bigg(-\lambda_{a} \log_2 \bigg(1+\frac{ P_{a,c,b} g_{a,c,b}}{ \sum\limits_{a=1}^A \sum\limits_{c=1}^C \sum\limits_{b=1}^B P'_{a,c,b}  f_{a,c,b}+\sigma_a^2} \bigg) \nonumber\\&+ \eta_b P_{a,c,b} + v P_{a,c,b} h_{a,c,b} \Bigg). \nonumber
\end{align}

By substituting the value of SINR, we get
\begin{align}
& 
= \frac{\partial}{\partial P_{a,c,b}} \Bigg(-\lambda_{a} \log_2 \bigg(1+\frac{P_{a,c,b} g_{a,c,b}}{\sum\limits_{a=1}^A \sum\limits_{c=1}^C \sum\limits_{b=1}^B P'_{a,c,b}  f_{a,c,b}+\sigma_a^2} \bigg)\nonumber\\& + \eta_b P_{a,c,b} + v P_{a,c,b} h_{a,c,b} \Bigg). \nonumber
\end{align}

\textcolor{black}{Applying the KKT conditions for optimality, and by solving the first derivative test, we obtain}
\begin{align}
= &-\lambda_a \Bigg(\frac{1}{1+\frac{P_{a,c,b}+g_{a,c,b}}{\sum\limits_{a=1}^A \sum\limits_{c=1}^C \sum\limits_{b=1}^B P'_{a,c,b}  f_{a,c,b}+\sigma_a^2} }\Bigg)\nonumber\\&\times \frac{\partial}{\partial P_{a,c,b}} \Bigg( 1+ \frac{P_{a,c,b} g_{a,c,b}}{ \sum\limits_{a=1}^A \sum\limits_{c=1}^C \sum\limits_{b=1}^B P'_{a,c,b}  f_{a,c,b}+\sigma_a^2} \Bigg)+ \eta_b + v h_{a,c,b}. \nonumber
\end{align}

Now, solving the partial derivative as
\begin{align}
& = \eta_b + v h_{a,c,b}-\lambda_a \nonumber \\&\times\Bigg( \frac{\sum\limits_{a=1}^A \sum\limits_{c=1}^C \sum\limits_{b=1}^B P'_{a,c,b}  f_{a,c,b}+\sigma_a^2 }{\Big(\sum\limits_{a=1}^A \sum\limits_{c=1}^C \sum\limits_{b=1}^B P'_{a,c,b}  f_{a,c,b}+\sigma_a^2 \Big)+ P_{a,c,b} g_{a,c,b}} \Bigg) \nonumber \\& \times\Bigg( \frac{ g_{a,c,b} \sum\limits_{a=1}^A \sum\limits_{c=1}^C \sum\limits_{b=1}^B P'_{a,c,b}  f_{a,c,b}+\sigma_a^2} {\Big(\sum\limits_{a=1}^A \sum\limits_{c=1}^C \sum\limits_{b=1}^B P'_{a,c,b}  f_{a,c,b}+\sigma_a^2 \Big)^2} \Bigg). \nonumber     
\end{align}

Taking L.C.M. of previous step and finding the partial derivative of internal function, we have
\begin{align}
&= -\lambda_a \Bigg(\frac{g_{a,c,b} \Big(\sum\limits_{a=1}^A \sum\limits_{c=1}^C \sum\limits_{b=1}^B P'_{a,c,b}  f_{a,c,b}+\sigma_a^2 \Big)^2}{ \Big(\sum\limits_{a=1}^A \sum\limits_{c=1}^C \sum\limits_{b=1}^B P'_{a,c,b}  f_{a,c,b}+\sigma_a^2 \Big)+ P_{a,c,b} g_{a,c,b}} \Bigg) \nonumber \\& \times\Bigg( \frac{1}{\Big(\sum\limits_{a=1}^A \sum\limits_{c=1}^C \sum\limits_{b=1}^B P'_{a,c,b}  f_{a,c,b}+\sigma_a^2\Big)^2} \Bigg)+ \eta_b + v h_{a,c,b}. \nonumber
\end{align}

Further simplification results in
\begin{align}
&  
= -\lambda_a \Bigg(\frac{g_{a,c,b}}{ \sum\limits_{a=1}^A \sum\limits_{c=1}^C \sum\limits_{b=1}^B P'_{a,c,b}  f_{a,c,b}+\sigma_a^2+ P_{a,c,b} g_{a,c,b}} \Bigg) \nonumber \\
& + \eta_b + v h_{a,c,b}.
\end{align}

By canceling the common terms and simplifying, we have
\begin{align}
&
 -\eta_b -v h_{a,c,b}=-\lambda_{a} \nonumber \\
& \Bigg( \frac {g_{a,c,b}}{ \Big(\sum\limits_{a=1}^A \sum\limits_{c=1}^C \sum\limits_{b=1}^B P'_{a,c,b} f_{a,c,b} + \sigma_a^2 \Big)+ P_{a,c,b} g_{a,c,b}}\Bigg) \nonumber 
 \end{align}
 \begin{align}
 &
\frac{1}{ \Big(\sum\limits_{a=1}^A \sum\limits_{c=1}^C \sum\limits_{b=1}^B P'_{a,c,b} f_{a,c,b} + \sigma_a^2 \Big)+ P_{a,c,b} g_{a,c,b}} = \frac{ \eta_b +v h_{a,c,b}}{\lambda_{a,c,b} g_{a,c,b}}. \nonumber 
\end{align}

After some mathematical simplifications, we get
\begin{align}
&
 \bigg(\sum\limits_{a=1}^A \sum\limits_{c=1}^C \sum\limits_{b=1}^B P'_{a,c,b} f_{a,c,b} + \sigma_a^2 \bigg)+ P_{a,c,b} g_{a,c,b}= \frac{\lambda_{a,c,b} g_{a,c,b}}{\eta_b +v h_{a,c,b}}. \nonumber 
 \end{align}

Inverting both sides of the equation as
 \begin{align}
 & 
P_{a,c,b} g_{a,c,b}= \Big(\frac{\lambda_{a,c,b} g_{a,c,b}}{ \eta_b +v h_{a,c,b}} \Big)- \Big(\sum\limits_{a=1}^A \sum\limits_{c=1}^C \sum\limits_{b=1}^B P'_{a,c,b} f_{a,c,b} + \sigma_a^2 \Big). \nonumber 
\end{align}

After cross multiplying, we have
\begin{align}
&
P_{a,c,b}= \frac{\lambda_{a,c,b} g_{a,c,b}-\bigg( \Xi \eta_b + v h_{a,c,b} \bigg)}{\eta_{b}+v h_{a,c,b}}. \nonumber
\end{align}
where $\Xi=\Big(\sum\limits_{a=1}^A \sum\limits_{c=1}^C \sum\limits_{b=1}^B P'_{a,c,b} f_{a,c,b} + \sigma_a^2 \Big)$.
\ifCLASSOPTIONcaptionsoff
  \newpage
\fi



\bibliographystyle{IEEEtran}
\bibliography{References}
%
\end{document}